\documentclass[aps,preprint,prl]{revtex4}
\newcommand{\zo}{z_{\alpha 1}}
\newcommand{\zt}{z_{\alpha 2}}
\newcommand{\kbo}{k_{B}}

\usepackage{amsfonts}
\usepackage{amsmath}
\usepackage{amssymb}
\usepackage{graphicx}

\begin{document}

\title{Maximum thermal conductivity of aligned single wall carbon nanotubes}
\author{A.L. Efros}
\affiliation{Department of Physics, University of Utah, Salt Lake
City UT, 84112 USA}
\date{\today}

\begin{abstract}
I estimate   maximum thermal conductivity $\kappa$ of a perfectly aligned bundle of  single wall carbon nanotubes. Each row of aligned nanotubes has a discrete structure. It consists of  segments of nanotubes with  length $L$. The spacing  between the segments block  the phonon path through the row.   Only the scattering due to the finite length of the segments  is taken into account. The result is that the 'effective'' mean free path is of the order of $L/7$. For 1 micron tubes (10,10) we get maximum value of $\kappa\approx 300$W/m K at room temperature. This result is in a reasonable 
 agreement with the experiment by Hone {\it et al.} assuming that in their samples $L\approx 1\mu{\rm m}$
\end{abstract}
\maketitle

\section{Introduction}
It is expected that  a single wall carbon nanotube(SWCN)is  a very promising object for creation of metamaterials  with a high thermal conductivity (TC)\cite{chush,chush3}. The first reason for this expectation is  that the carbon-based materials, like diamond, have the largest known TC and the second reason is   a molecular perfection\cite{chush} of the (SWCN's). However, to the best of my knowledge, the highest  TC ever observed in SWCN's bundles at room temperature is about 220 W/mK and it is ten times smaller than  TC of  the natural diamond\cite{antony}.  This highest result has been reported by Hone {\it et al.}\cite{hone} for a bulk sample of magnetically aligned nanotubes. The aligned SWCN's form a bundle where all tubes have a preferable orientation in  some direction.  Hone {\it et al.}\cite{hone} show that the TC of the  aligned SWCN's is strongly anisotropic with the largest value in the direction of the alignment.
The enhancement of the TC due to the alignment has been observed also by Zhou {\it et al.}\cite{Zhou} and by Choi {\it et al.}\cite{Choi}, but the absolute values of the reported TC have been significantly smaller than in Ref.\cite{hone}.
There are many theoretical works on  TC of the SWCN's. Some computational
 ones\cite{berber,maru,osman,goddard} are made by molecular
dynamics simulations. The results of these simulations have
different values and different $T$-dependences. They predict mostly very high values of the TC. 
  We think  that the main problem 
of  all these works is a small  size of an array that can be simulated. There are  also some different analytical approaches to the problem\cite{zheng,cao} and beautiful
reviews\cite{book2,nanoreview,anvar}.

The purpose of my  work is to estimate the maximum TC  value  of 
aligned nanotubes taking into account that they consist of segments with a finite length. It is well known, that tubes in ropes are not infinitely long, but have brakes, because each  method of synthesis
is able to create separated tubes of only a certain length. It is believed that this length is of the order of a few $\mu$m (See
Ref.\cite{Yos} and references therein)
Then due to Van der Waals forces the tubes stick together and create bundles where the end of tube has no chances to make a  strong chemical bond to the end of neighboring tube.
There are many experiments that show, that tubes inside bundles have free ends. The idea of my work is to argue that this effect may be responsible for the relatively low TC as compared to the  crystalline carbon materials.

 I consider  a bundle of nanotubes perfectly aligned in $x$ direction. Each segment of a nanotube
has a finite   length with an average value  $L$. The nanotubes are organized in an ideal
triangular lattice with 6 nearest neighbors\cite{phononreview}. The cross section in a
plane perpendicular to the nanotubes is shown in Fig. \ref{fig123}
(a). The cuts in each line of the nanotubes have  random positions. Thus, on the
length of each segment there are in average six cuts of its nearest
neighbors. A homogeneous  interaction between  infinite tubes  
 does not cause the loss of the
phonon momentum. However, a  phonon flux has to overcome the openings between the  segments at the termination points of each nanotube segment. I assume that this openings are so large that  a jump  of a flux  occurs with an assistance of all six neighboring rows of the tubes as it shown in  Fig. \ref{fig123} (b). Slightly  different mechanism   of momentum  scattering appears in a given nanotube (``0'') if one of  the neighboring nanotubes has a termination point as shown in Fig. \ref{fig123}(c). 

 Having in mind to get a maximum estimate of the TC, the   propagation of  heat flux $Q$ between the scattering points I assume to  be ballistic. Quick phonon exchange at the  scattering points leads to a thermalization of symmetrical parts of  distribution functions of phonons with  temperatures that are determined by  values of effective  thermal resistances between scattering points.
It should be noted that my calculations cannot be applied directly to multiwall carbon
nanotubes. 
\section{Scattering problems}
For simplicity  an elastic scattering only is considered. This means  that
 phonons generated in  neighboring nanotubes  not only have the same frequency, but belong to the same mode as the  incident wave. It is not necessarily true but it 
should not lead to a big mistake  due to small matrix element of the inelastic scattering. On the other hand, without this approximation the calculation of the TC would be a mess because one cannot  consider each mode independently.  I  assume a linear dispersion law for the modes under
study and argue that it is also not important if the interaction is
large. 

\begin{figure}
\includegraphics[scale=0.55]{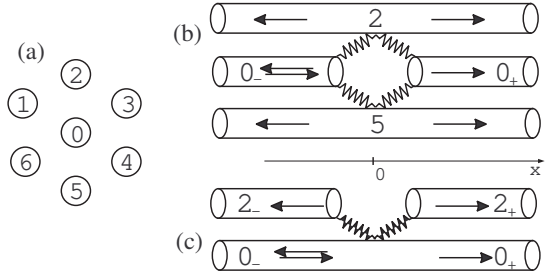}
\caption{\label{fig123}(a) The cross-section of the bundle that show
nanotube ``0'' and its nearest neighbors. (b) The first scattering
problem. The cross-section by the plane of nanotubes  2-0-5. The wave incident from
$0_-$ reflects backward and transmits through the opening  into $0_+$
with simultaneous excitation  of the waves in all six neighboring
tubes. (c) The second scattering problem. The wave incident from
$0_-$ and is scattered by the cut in tube 2. It reflects backward,
transmits into  $0_+$ and excites waves in  tube $2_+$.}
\end{figure}

Consider a  propagation of some  mode along the
nanotube  ``0'' as shown  in Fig. \ref{fig123} (a). There are two
basic scattering geometries for this mode. One of them appears at  the
terminational points of this very nanotube. It is shown in Fig.
\ref{fig123} (b).  Only two neighboring nanotubes are shown, however all six are
participating in the scattering. The interaction between the neighboring 
nanotubes   is shown by springs in
Fig. \ref{fig123}. The energy of this interaction is expected to be
larger than Van der Waals interaction  because $\mathrm{div} \mathbf{P}$ at the termination
point might be non-zero, where $\mathbf{P}$ is a polarization in the
wave. The second scattering
geometry (Fig. \ref{fig123}(c)) appears if one of the nearest
neighbors of nanotube 0 has a terminational point. In both cases only  an elastic interaction is considered.

In the first scattering problem the acoustic  wave  in the nanotube 0 is described by a following equation

\begin{eqnarray}
{\cal L}(U_{\mp})=K\delta(x)\sum_{i=1}^{6}(U_{i}-U_{\mp}),
\label{wave1}\\
{\cal L}(U)\equiv \frac{1}{s^{2}}\frac{\partial^{2} U}{\partial
t^{2}}- \frac{\partial^{2} U}{\partial x^{2}}, \label{Lop}
\end{eqnarray}
where $s$ is  velocity of the mode, $K$ is a characteristic of the
springs, $i=1 \ldots 6$, $U_{\mp}$ is the displacement of a tube ``0'' at $x<-0$ and at $x>+0$ respectively, $U_i$ are displacement of the neighboring tubes. (See Fig. \ref{fig123}b).

 The equations for the neighboring
tubes, coupled with  tube ``0'', are
\begin{equation}
{\cal L}(U_i)=K\delta(x)[(U_+ -U_i)+(U_- -U_i)].
\label{om}
\end{equation}

One can see  from the symmetry of equations that all $U_i$ are the
same. In what follows I put $U_i\equiv U_1$. The solution can be
found in a form
\begin{eqnarray}
U_-=e^{ i kx} +De^{-ikx} \ (x<0), \ \ U_+=C e^{ ikx} \ (x>0), \nonumber \\
U_1^-=Ae^{ -ikx} \  (x<0), \ \   U_1^+=A e^{ikx} \ (x>0).
\label{sol1}
\end{eqnarray}
The time dependent factor $\exp (-i\omega t)$ is omitted here and below.
Since the forces acting on the tubes are localized near the openings and the openings are supposed to be less than the wavelength, the
displacements have discontinuities at point $x=0$.
The boundary conditions can be obtained by integrating
Eqs.(\ref{wave1}-\ref{om}) over x. One gets that at $x=0$
\begin{eqnarray}
\frac{\partial U_1^-}{ \partial x}-\frac{\partial U_1^+}{ \partial
x}=
T(U_++U_--2U_1),\\
\pm\frac{\partial U_{\mp}}{ \partial x}=6K(U_1-U_{\mp}).
\label{bound}
\end{eqnarray}
The solution has a form
\begin{eqnarray}
|D|^2=\frac{k^4+73K^2k^2+36^2K^4}{ (36K^2+k^2)(49K^2+k^2)} \nonumber \\
\label{D}
|C|^2=\frac{K^4}{ (49 K^2+k^2)(K^2+k^2/36)} \nonumber\\
\label{C} |A|^2=\frac{36 K^2}{(42^2K^2+36k^2)}\nonumber.
\end{eqnarray}
Due to the energy conservation
\begin{equation}
12A^2+C^2+D^2=1.
\label{con}
\end{equation}
In the case of weak interaction $k/K\rightarrow \infty$ one gets
\begin{equation}
|D|^2=1, |C|^2=|A|^2=0,
\label{weak}
\end{equation}
 which means that the wave completely
reflects from the terminational point. In the opposite case of a
strong interaction  $k/K\rightarrow 0$,
\begin{equation}
|D|^2=36/49,~|C|^2=|A|^2=1/49.
\label{strong}
\end{equation}
 It will be 
important below that in the case of the strong interaction the structure of the operator $\cal L$ is irrelevant since I ignore the spatial derivatives in the boundary
conditions Eq.~(\ref{bound}). Therefore it does not matter whether or not this wave is acoustic and  what 
spectrum it has.  In fact, the numbers in Eq.~(\ref{strong}) is determined only  by
the geometry of the problem.

For the geometry of Fig.~\ref{fig123} (c), the equations of waves are
\begin{eqnarray}
{\cal L}(U)=K\delta(x)(U_2^-+U_2^+-2U),\\
{\cal L}(U_2^{\mp})=K\delta(x)(U-U_2^{\mp}),
\end{eqnarray}
where $U$ is displacement of tube 0,  $U_2^+ $ and $U_2^- $ are displacements of tube 2 at $x>0$ and $x<0$ respectively. The solutions have a form
\begin{eqnarray}
U^{-}=e^{i kx} +D_{1}e^{ -ikx}, \ \ U^{+}=C_{1}e^{ikx}, \\
U_2^{-}=A_{1}e^{-ikx}, \ \ U_2^{+}=A_{1}e^{ikx}.
\end{eqnarray}
The boundary conditions at $x=0$ are
\begin{eqnarray}
\frac{\partial U^-}{ \partial x}-\frac{\partial U^+}{ \partial x}=K(U_2^- +U_2^+ -
2U), \\
\pm\frac{\partial U_2^{\mp}}{ \partial x}=K(U-U_2^{\mp}).
\end{eqnarray}
Then
\begin{equation}
|D_1|^2=|A_1|^2=\frac{K^2}{ 4K^2+k^2}, \ \ |C_1|^2=\frac{K^2+k^2}{
4K^2+k^2}.
\label{tot}
\end{equation}
Due to the energy conservation $2|A_1|^2+|D_1|^2+|C_1|^2=1$. For the
weak interaction $K\ll k$ one gets
\begin{equation}
 |C_1|^2=1, 
|A_1|^2=|D_1|^2=0.
\label{1weak}
\end{equation}

 In
this case the wave  is transmitted without any scattering. In the opposite case
$K  \gg k$ one gets
\begin{equation}
|D_1|^2=|A_1|^2=|C_1|^2=1/4.
\label{1strong}
\end{equation}
As in the previous case the  result  for the strong interaction  is   independent of the form of the
operator $\cal L$.

\section{ Thermal Conductivity}

Let us now calculate the TC of the perfectly aligned nanotubes. In the approximation of 
elastic scattering 
 the heat flux $Q$ along each row of the aligned nanotube conserves because 
 the waves generated in neighboring nanotubes due to scattering
have zero total momentum. This leads to a conservation of $Q$ along the row because in the theory of
phonon thermal conductivity any relaxation of $Q$  is the result of  momentum loss. It is important,
however, that at the points
of scattering of both types, considered above, the numbers  of phonons in each mode changes. Therefore the symmetric parts of the distribution functions in these points can be considered  as in equilibrium with different temperatures for each point.   
 Finally, I assume that between the scattering
points of both types the propagation is ballistic.

\begin{figure}
\includegraphics[scale=0.5]{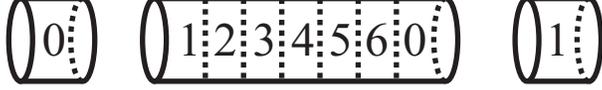}
\caption{\label{fig2}Part of the  row of nanotubes with two cuts.
Dotted lines in the nanotubes correspond to the cuts of the
neighboring nanotubes. The regions  between them are considered as
ballistic.}
\end{figure}

 In average
every section of a nanotube can be divided into seven ballistic regions such that
each  boundary   of the region corresponds to a cut  in  one
of the six neighboring rows of the nanotubes. The part of one row is
shown in  Fig.~\ref{fig2}. Since the energy flux is the same along the row, but
the scattering is different the temperature intervals between
neighboring boundaries  are also different. To calculate the TC, I find
the total temperature difference through all the nanotube at a given
flux $Q$.

Consider a region $i$  of one nanotube  and assume that each end 
of a region perfectly matches a thermal bath. The temperature difference of the left and right boundaries of the region is $t_i$. Thermal flux produced  in this region is $Q=G(T)t_i$, where the 
 function $G(T)$ is called thermal conductance. It can be written in a form\cite{expquant}  
\begin{equation}\label{g2}
G (T) = \frac{k_B^2 T}{h} \sum_{\alpha} \int_{\zo}^{\zt} dx \
\frac{x^{2}\exp(x)}{(\exp(x)-1)^{2}},
\end{equation}
where
 $z=\hbar \omega/\kbo T$, and the sum
is over all monotonously increasing segments of spectrum
$\omega_{s}(k)$, $\zo$ and $\zt$ being the lower and upper boundaries
of such segments. Here $k_{B}$ and $h$   are the Boltzmann and the
 Plank constants respectively. To calculate the above integrals one should know the vibration
spectra of nanotubes. They  have been calculated previously within
different frameworks such as an empirical force constant
model\cite{book1,phononreview}, \textit{ab initio}
studies\cite{abini1}, and tight-binding molecular dynamics\cite{tbmd}. I  use here the 
function $G(T)$ calculated by Yu. Gartstein\cite{gart}. His results for phonon spectra does not differ much from 
the previous ones though he used some original method.

Suppose that the temperature decreases from left to right and the
flux oriented in the same direction. The flux coming from the section
0 to section 1 is $Q|C|^2$, where $|C|^2$ is   the transmission coefficient of
the first scattering problem.
  Ballistic
flux generated  in the region 1 is $-Gt_1$, where $G$ is ballistic conductance
given by Eq. (\ref{g2}) and $t_1$ is the negative temperature
difference between boundaries 0-1 and 1-2 in  Fig.2. To get the total flux in the region 1
one should take into account reflection at the boundary 1-2. Simple calculations show  that multiple reflections inside 
the same region do not change the result substantially.
Finally, in the region 1
\begin{equation}
Q=(Q|C|^2-Gt_1)(1-|D_1|^2).
\end{equation}
Thus I can find the change of the temperature $t_1$ in the first region
\begin{equation}
-t_1={Q\over G}\left({1\over 1-|D_1|^2}-|C|^2\right).
\label{1}
\end{equation}
In the same way I find
\begin{equation}
-t_2={Q\over G}\left({1\over 1-|D_1|^2}-|C_1|^2\right)=-t_3=...=-t_6,
\label{26}
\end{equation}
while
\begin{equation}
-t_0={Q\over G}\left({1\over 1-|D|^2}-|C_1|^2\right).
\label{0}
\end{equation}
The total change of the temperature $\Delta T$ through the segment  can be calculated as
\begin{equation}
\Delta T=\sum_{i=0}^6t_i=-Q B/G(T),
\label{Q}
\end{equation}
where
\begin{equation}
B=\left({6\over  1-|D_1|^2}+{1\over  1-|D|^2}-6|C_1|^2-|C|^2\right).
\label{B}
\end{equation}
 Since I  suppose that the interaction
is strong, coefficients in $B$ are independent of the frequency. As a result I
 can express the TC $\kappa$ through $G(T)$.

Flux per ${\rm m}^2$ is $Q_T=QN$, where $N$ is the number of tubes in the bundle per ${\rm m}^2$. Assuming triangular lattice it is easy to get
  $N=4\times 10^{17}$m$^{-2}$  for (10,10) nanotubes\cite{book2,phononreview}. 
Using equation   $\Delta T= L dT/dx $, definition of TC  $Q_T=-\kappa dT/dx$,
and Eq.(\ref{Q}) one gets
\begin{equation}
\kappa=N LG(T)/B.
\label{ka}
\end{equation}

One can see from Eq.(\ref{weak})and Eq.(\ref{1weak}) that in the case of the weak interaction $B\rightarrow \infty$ and
$\kappa\rightarrow 0$. 
That is because the jump of the wave through the cut is a bottle neck of the problem. Thus, assuming that interaction is strong, I   get a maximum estimate for the TC. 

From Eqs.~(\ref{strong},\ref{1strong}) one can find that $1/B=0.0976$, and Eq.(\ref{ka}) leads to the final result 
\begin{equation}
 \kappa=0.0976 G(T) L N.
\label{final}
\end{equation}
where $L$ is the length of one nanotube.

\begin{figure}
\includegraphics[width=7.5cm]{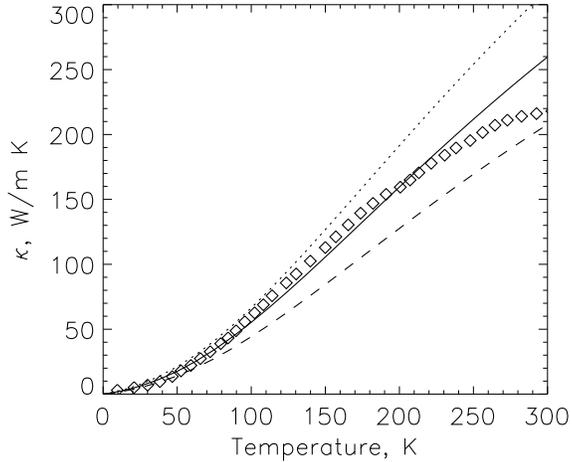}
\caption{\label{fig5}Thermal conductivity as calculated from
Eq.~(\ref{final}) for the (10,10) tube with $L=0.7$ (dashed line), 0.87
(solid) and 1.0 (dotted) micron. The experimental data of
Ref.~\onlinecite{hone} are shown by diamonds.}
\end{figure}

Fig. \ref{fig5} shows the results of Eq.(\ref{final}) at $L=1,0.87, 0.7\mu$
together with the results by Hone {\it at al.}\cite{hone}. One can
see that the theory reflects well enough both the magnitude and the
temperature behavior. In fact, the only parameter here is the average
length of a nanotube. The deviation at
high temperatures is probably related to Umklapp processes.

\section{Conclusions}
Finally, I present the maximum estimate of the TC of perfectly aligned nanotubes taking into account  the  scattering of phonons by the
terminational points of the nanotubes. This estimate  gives a quantitatively correct
description of the thermal conductivity of aligned nanotubes as obtained experimentally by Hone {\it et al.}\cite{hone} assuming that the length of segments is of the order of 1 $\mu$m. It follows from my  results that the  way to make 
thermal conductivity of the aligned nanotubes at room temperature larger than about 300 W/m K is to increase their lengths. Of course 
the TC will not increase indefinitely with $L$, as it follows from  Eq.(\ref{final}), because sooner or later the mean free path due to other scattering processes will  be  smaller than $L/7$.  However, some additional gain may be achieved with increasing $L$.

I am grateful to  V.  Agranovich and A.  Zakhidov  for fruitful
discussions. The work has been funded by DARPA subcontract SC0302

\bibliographystyle{apsrev}
\bibliography{pebib}
\end{document}